# Strengthening of Mg-Al-Ca alloys with C15 and C36 Laves phases

M. Zubair[1,2,*], Stefanie Sandlöbes-Haut[1], Risheng Pei[1], Maximilian A. Wollenweber[1], James S.K-L. Gibson[1], S. Korte-Kerzel[1,*]

[1]Institute for Physical Metallurgy and Materials Physics, Kopernikusstr. 14, RWTH Aachen University, 52074 Aachen, Germany.

[2]Department of Metallurgical and Materials Engineering, G.T Road, UET Lahore, Pakistan.

*Corresponding authors zubair@uet.edu.pk, korte-kerzel@imm.rwth-aachen.de

## Abstract

The Laves phase skeleton in cast Mg-Al-Ca alloys is known to provide considerable strengthening. Laves phases such as $CaMg_2$ (C14), $Ca(Al,Mg)_2$ (C36), and $CaAl_2$ (C15) have high melting points, high hardness at room and elevated temperatures, but unfortunately are inherently brittle. Mg-Al-Ca alloys thus have good creep properties but limited ductility. An understanding of the co-deformation behaviour of α-Mg and Laves phases is essential for optimising the strength-ductility balance of these alloys. Here, we study the mechanical behaviour of a Mg-4.65Al-2.82Ca alloy using micropillar compression in the α-Mg matrix, at α-Mg/C36 and α-Mg/C15 interfaces and in the C15 phase in combination with scanning electron microscopy (SE imaging), electron backscatter diffraction (EBSD), transmission Kikuchi diffraction (TKD), and low-kV scanning transmission electron microscopy (STEM). We show that both, C15 and C36, Laves phases provide considerable strengthening to the α-Mg matrix by delaying the onset of basal slip and extension twinning, while only the C36 phase appears to allow a certain extent of slip transfer/ plastic co-deformation, in spite of its greater anisotropy compared with the cubic C15 phase. We therefore conclude based on these results that strengthening of the α-Mg matrix by the C36 Laves phase is preferable given that it combines easy skeleton formation with some co-deformation and considerable stability at common application temperatures of magnesium alloys.

# 1 Introduction

Laves phase containing Mg-Al-Ca alloys exhibit superior creep resistance when compared to conventional Mg-Al-Mn or Mg-Al-Zn alloys [1-3]. These Laves phases reinforcing the Mg-Al-Ca alloys are characterised by high strength [4], hardness [4-7] and melting points [5] when compared to the α-Mg matrix [6]. In cast Mg-Al-Ca alloys, the Laves phases are present as an intra- and intergranular skeleton within the α-Mg matrix [6, 8-11]. Knowledge about the amount of strengthening induced by Laves phases in the Mg matrix is important and can be helpful for alloy design. However, even where overall strengthening of an alloy is observed due to the introduction of a second phase, it is the understanding of which deformation mechanisms are affected, and in which way, that really opens purposeful microstructure and composition design strategies.

Previous work on Mg-Al-Ca alloys has mainly focussed on the identification of advantageous microstructures with respect to the formation of an intermetallic skeleton and the macroscopic properties that can be obtained as a result [2, 8, 11-13]. For those alloys with the most promising properties, the co-deformation mechanisms were also explored [6, 10]. These alloys have in common that they contain the $Ca(Mg,Al)_2$ C36 phase as the dominant strengthening Laves phase. What is not known is whether one of the other Laves phases of the Mg-Al-Ca system may in principle give better co-deformation performance in offering equally or better high temperature stability and creep resistance while allowing plastic co-deformation of the matrix and intermetallic skeleton at stress concentrations to avoid cracking and void formation that lead to early failure. In this respect, the cubic $CaAl_2$ phase is of particular interest. It is known to be harder than the other Laves phases [5, 14, 15], making it possibly harder to co-deform, but as a cubic phase also likely offers a larger number of slip systems, which may align with the basal plane of the hexagonal Mg matrix to allow dislocations to cross the interface.

The most easily activated deformation mechanisms of Mg at room temperature are basal slip and extension twinning. This is because the critical resolved shear stresses, $\tau_{CRSS}$, required to activate basal slip and extension twinning, amount to ≈ 0.5 MPa [16, 17] and <10 MPa [18, 19], respectively. On the other hand, the stresses required to activate prismatic and pyramidal <a> slip in Mg are ≈ 39-44 MPa [20-23]. This large difference between the $\tau_{CRSS}$ for activation of different deformation mechanisms usually results in a strong mechanical anisotropy at ambient temperature [24, 25]. Consequently, large differences are found between the yield stresses when the crystals are deformed along different directions [26-28]. Considering this, if micropillar testing is carried out on different crystal orientations in a comparison of micropillars from the two different microstructural regions (α-Mg and α-Mg/Laves phase interfaces), then the orientation dependence of deformation in the α-Mg matrix has to be taken into account as well for any specimens consisting of both phases, particularly as deformation is bound to start in the softer metallic phase.

Depending on the Ca/Al ratio and the temperature chosen for any annealing treatment, there can be three different types of Laves phases in Mg-Al-Ca alloys: $CaAl_2$ (C15), $CaMg_2$ (C14) or $Ca(Mg,Al)_2$ (C36) phase [8, 14, 29-32]. The C36 Laves phase is present in as-cast alloys with intermediate Ca/Al ratio (≈ 0.6) and similar compositions like the one studied in this work [8-11, 33, 34].

Although they are macroscopically brittle at low temperatures, Laves phases have shown considerable plasticity in small scale mechanical testing like micropillar compression [4, 7, 15, 35]. The size of Laves phase struts in conventional as-cast Mg-Al-Ca alloys is of the order of 1 μm or less; close to the dimensions generally studied in micropillar compression testing.

Here, our aim is to achieve insights into the co-deformation of the Mg matrix with two different Laves phases under conditions close to those in a final, technologically useful microstructure. Due to the intrinsic geometry of the Laves phase interfaces available from as-cast microstructures, with their finer Laves precipitates, certain information will remain out of reach, e.g. an accurate interfacial shear strength or quantitative transmission stresses. However, even though the occurring phase transformations make the preparation of interfaces for dedicated mechanical testing a challenging endeavour, this information may be studied separately in the future if straight interfaces can be achieved by appropriate heat treatment or solidification conditions.

In this work, we therefore set out to study the active deformation mechanisms and in particular co-deformation in a Laves phase reinforced Mg-Al-Ca alloy. For this, we considered four systematic sets of conditions: (1) (micro)compression of the α-Mg phase in different orientations to assess the behaviour of the matrix alloyed with Al and Ca as a baseline, (2) compression of micropillars containing α-Mg/Laves phases interfaces with the C36 Laves phase from as-cast alloy or (3) Mg/C15 Laves phase interfaces from the same alloy in an annealed state and (4) compression of C15 Laves phase. Together, these allow us to directly compare the strengthening effect of the Laves phases on the α-Mg matrix and assess the effects of the geometrical alignment of Laves phases on the co-deformation behaviour.

# 2 Experimental methods

## 2.1 Sample preparation and microscopy

The raw materials were molten in a steel crucible inside a vacuum induction furnace under a protective atmosphere of Ar. The melt was solidified and cooled in a Cu mould inside the furnace under protective atmosphere. The chemical composition of the as-cast alloy was measured using wet chemical analysis to be Mg-4.65Al-2.82Ca alloy. To increase the size of the intermetallic struts, a heat treatment of the as-cast alloy was carried out at 500 °C for 48 hrs in Ar atmosphere followed by cooling in air. This resulted not only in the coarsening of the intermetallic structure but also in the transformation from the C36 to the C15 phase. We refer to these two states as the as-cast and annealed conditions.

Two specimens for microstructure analysis and micropillar milling were cut using electric discharge erosion from the as-cast block, one of which was subjected to the heat treatment. The samples were ground using 2000 and 4000 SiC emery papers, followed by mechanical polishing using 3 and 1 μm diamond suspension. To remove the deformation layer, the samples were subjected to electro-polishing using Struers AC-II electrolyte. The samples were etched for 60 s at ~-20 °C and 15V. The electro-polished surface was then subjected to mechanical polishing using Struers OPU (colloidal suspension of $SiO_2$ nano particles) for ~60 s to remove the waviness arising from electro-polishing of the dual phase material.

Secondary electron (SE) imaging in scanning electron microscopes (SEM, FEI Helios 600i and Zeiss LEO1530) was used to analyse the microstructures. Energy dispersive spectroscopy (EDS) in SEM was used to determine the composition of the α-Mg matrix and the Laves phase. Electron backscatter diffraction (EBSD) was used to measure the crystallographic orientations of the α-Mg matrix. Due to the small size of the intermetallic Laves phase in the as-cast alloy, it was not possible to measure the orientation of the Laves phase. An accelerating voltage of 10 kV was used for SE imaging and EDS while EBSD was done at 20 kV.

Transmission Kikuchi Diffraction (TKD) was done on the cross-sections of two selected micropillars (see section 2.2). For this purpose, the pillars' cross-sections were lifted out of the original sample and were placed on a TEM grid. They were thinned down to a thickness of ~300 nm using focussed ion beam (FIB) with currents ranging from 0.230 nA to 80 pA. A small current, i.e, 80pA, was used for the final polishing steps. TKD was performed on the cross-sections with a step size of 50-60 nm at 30 kV and 5.5 nA. All EBSD data was analysed using OIM Analysis (EDAX).

## 2.2 Microcompression

Micropillars of ~1.6–2 µm diameter and ~ 3.5–4.5 µm height were milled in matrix grains with different orientations, where the inclination of the c-axis to the compression direction was varied from 0 to 90 °, using FIB milling. Stress and strain calculations were performed using the actual dimensions of the pillars determined from SEM images. The top diameter of the pillars was used for stress calculation. The cylindrical pillars were milled using concentric circles with a decreasing ion beam current from 9.3 nA to 80 pA at an accelerating voltage of 30 kV. The pillar taper angle was less than 2° for most pillars.

The micropillars were compressed in two different load-controlled nanoindenters (iNano and InSEM III; Nanomechanics Inc., USA) at a loading rate of 0.05 mN/s and using a flat punch diamond indenter with a diameter of 10 µm. The Sneddon correction was applied to the depth data to minimise the effect of elastic deformation of substrate and indenter as discussed in [36]. The tests were stopped after a significant strain burst was observed in the load-depth curve. The slip systems were analysed using the code and method already presented in [37], after adjustment for the hexagonal Mg phase. The micropillars before and after deformation were imaged from three different directions at a tilt angle of 45 °.

# 3 Results

## 3.1 Microstructural characterisation

A typical SE image of the as-cast microstructure of cast Mg-4.65Al-2.82Ca is presented in Figure 1 (a). The α-Mg matrix (grey colour) is reinforced with an interconnected intermetallic skeleton (white phase) shown at greater magnification in Figure 1 (b). The grain size of the alloy is big when compared to the areas bounded by the intermetallic struts (Figure 1 (c)). The big grains enabled milling of micropillars in each grain within the α-Mg matrix and at the α-Mg/Laves phase interfaces. The determined composition of the intermetallic Laves phase and the α-Mg matrix is presented in Figure 1 (d). Note that the solubility of Ca is very low and therefore not quantified here by means of EDS. The morphology and composition of the Laves phase suggest that it is the $Ca(Mg,Al)_2$ C36 phase. The C36 phase was also determined to be the main intermetallic phase in similar alloys studied previously [9-11, 34]. This phase is also shown to contain small precipitates as highlighted by green arrows (Figure 1 e). These precipitates may strengthen or toughen the C36 Laves phase depending upon its composition. Moreover, there are intragranular needle-like $CaAl_2$ precipitates within the α-Mg matrix in line with the previously reported work on similar alloys [8, 11].

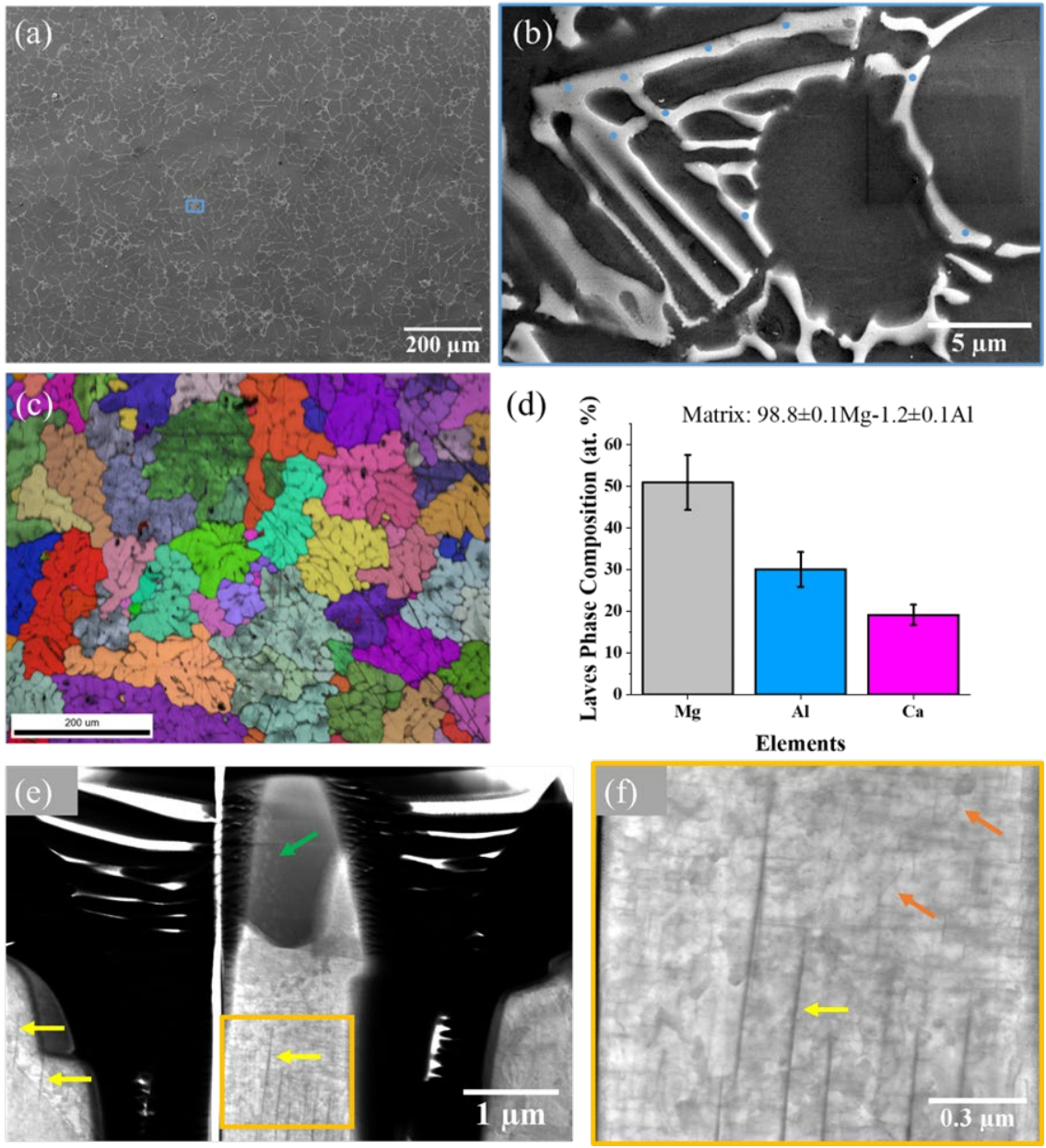

*Figure 1: (a) SE image showing the microstructure of the as-cast Mg-4.65Al-2.82Ca alloy, (b) is the high magnification image of the microstructural region highlighted by blue rectangle in (a), (c) shows the inverse pole figure (IPF) and image quality (IQ) maps superimposed on each other showing that the grain size is bigger than the distance between intermetallic struts (IPF legend is same for Mg phase as presented in Figure 2) and (d) depicts the composition of the microstructural constituents determined using EDS. The blue spots in (b) highlight the locations of the EDS spots made on the intermetallic Laves phase. (e) STEM image of a cross-section through a micropillar from the as-cast material containing C36 Laves phase as the darker phase with precipitates (green arrow) and α-Mg matrix with dislocations (orange arrows) and $CaAl_2$ precipitates (yellow arrows) including a magnified view of the orange rectangle.*

The microstructure of the heat-treated alloy contains the α-Mg matrix and inter- and intragranular $CaAl_2$ Laves phase precipitates (Figure 2). The $CaAl_2$ phase precipitates were significantly different in shape compared with the C36 Laves phase present in as-cast alloy where they were present as round particles or thick platelets in comparison to thin interconnected struts in the as-cast alloy, Figure 2 (a and b). Moreover, the composition of the intermetallic phase drastically changed after heat treatment: in the as-cast alloy the C36 Laves phase was observed, while in the heat-treated sample the $CaAl_2$ phase was observed (Figure 2 c). The larger, rounder, and more homogeneously distributed Laves phase precipitates in the heat-treated alloy enabled us to determine the orientation of both phases. The $CaAl_2$ Laves phase precipitates share no specific orientation relationship with the α-Mg matrix (Figure 2 d and e). The area fraction of the Laves phase is reduced from ~ 6.5 % in the as-cast state (determined from the micrograph shown in Figure 2 (a)) to ~ 4.9 % in the heat-treated state (determined from the micrograph given in Figure 2 (b)). This reduction in area fraction is assumed to occur because the C36 phase contains a significant amount of Mg while the C15 phase can accommodate only a much smaller amount of Mg in solid solution [38].

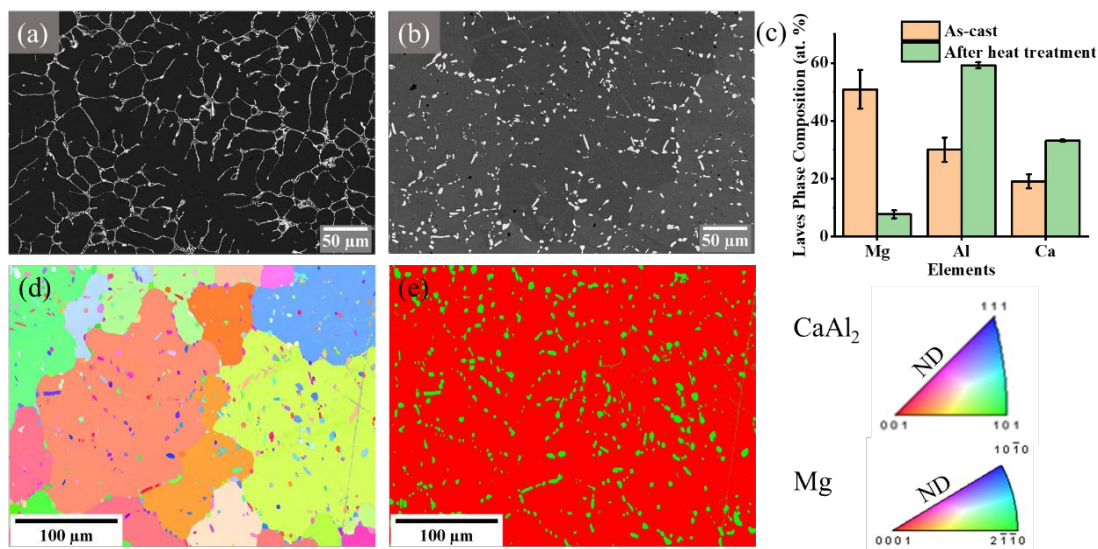

*Figure 2: (a) Microstructure of the as-cast alloy, (b): microstructure of the heat-treated alloy, (c): change in composition of the intermetallic phase after heat treatment, (d): IPF map of Mg and $CaAl_2$ phase depicting random orientation relationship between matrix and the second phase and (e): phase map of the same region shown in (d) having Mg (in red) and $CaAl_2$ in green. ND is the surface normal direction pointing towards the reader.*

## 3.2 Deformation of the α-Mg phase – effect of c-axis orientation

The micropillars were milled into α-Mg grains with declination angles (ϕ) between the compression axis and the c-axis of the Mg unit cell varying from ~6 to ~89 °. As expected for Mg, the engineering stress-strain curves showed a strong dependence on the orientation (Figure 3 a). The yield stress first decreased up to intermediate angles (see also Figure 3 a) and then increased again towards 89 ° (Figure 3 b). Basal slip was found to be the predominant deformation mechanism at the angles tested with ϕ up to 55° (the next considered inclination was 77°, all tested angles are shown in Figure 3 b). The slip system identification was confirmed by matching the deformed pillars with simulated pillars which were generated using the Matlab code and EBSD data as discussed in [37, 39]. For this purpose, the deformed and simulated pillars were compared from three different sides and rotated clockwise at increments of 120 ° (Figure 3 c). The deformation features in micropillars undergoing basal slip were very similar to the earlier reported cylindrical Mg micropillars oriented for basal slip [40-42].

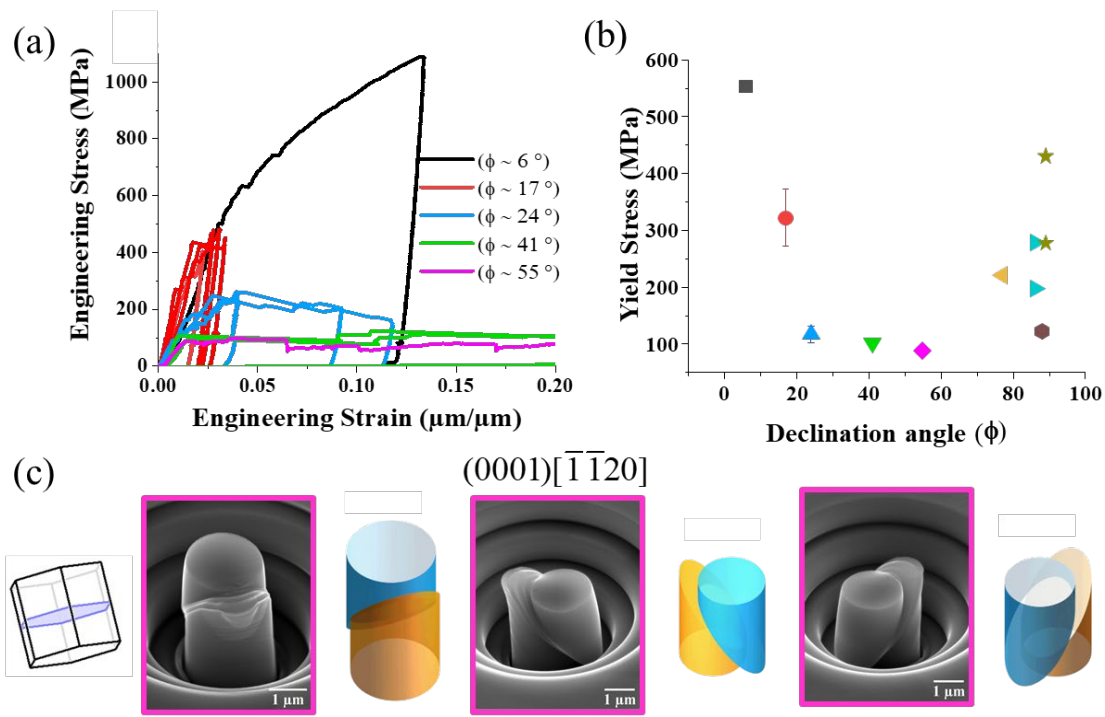

*Figure 3: (a) Engineering stress-strain curves of Mg micropillars in different orientations deformed at room temperature for low to intermediate inclination angles ϕ, (b): variation of the yield stress ($\sigma_{0.2}$) with ϕ, (c): SE images of the side view taken at an angle of 45° to the surface normal of a deformed micropillar from three different directions and their comparison with simulated micropillars depicting the activation of basal slip,*

Extension twinning or a combination of extension twinning and basal slip (within the deformation twin) was observed when ϕ was varied between 77 and 89 ° (Figure 4). In a few micropillars, only basal slip was also observed together with bending of the micropillars (one such example is shown in Figure 4 (e)). A side and top view image of one deformed micropillar with the c-axis nearly perpendicular to the applied load is presented in Figure 4 (a) and (b). As opposed to basal slip, it is not straight forward to confirm extension twinning in the deformed micropillars. We therefore used post-deformation orientation analysis by TKD to confirm the occurrence of twinning by lifting out a thin lamella. The position from where the lamella was lifted out is highlighted by the white dotted rectangle in Figure 4 (b). The IPF map of the deformed cross-section (Figure 4 c and d) reveals that the entire micropillar has twinned and analysis of the misorientation angle between the undeformed matrix and the deformed pillar confirms that the twin is a $\{10\bar{1}2\}$ extension twin commonly observed in Mg.

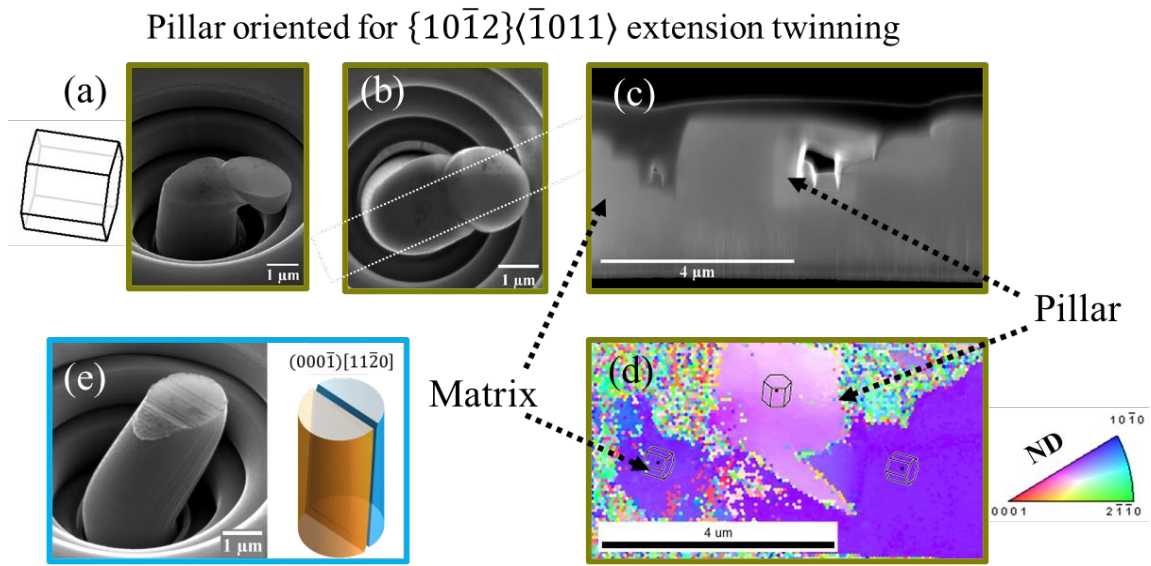

*Figure 4. (a): side view (SE image) of a deformed micropillar oriented for extension twinning (unit cell and micropillar image is at 45° tilt), (b): top view of the micropillar shown in (d). The white dotted rectangle in (b) depicts the region from where the FIB lamella was lifted out for TKD, (c): micropillar cross-section on which TKD was performed, (d): IPF map revealing that the entire pillar has twinned, (e): side view of a micropillar which underwent basal slip even when the highest SF for basal slip was only 0.06 (ϕ = 86°).*

### 3.3 Co-deformation of the α-Mg(Al,Ca) and hexagonal C36 Laves phase

In micropillars containing both phases in the as-cast alloy, i.e. Mg matrix and mainly C36 Laves phase, the Laves phase skeleton can delay the onset of basal slip and twinning, but the effectiveness of this strengthening effect depends on the geometric alignment of intermetallic phase and the active deformation paths originating from the Mg matrix (Figure 5).

There is a significant increase in yield stress when the Laves phase runs through the micropillar (Figure 5 a-c). The strengthening effect of the Laves phase was much weaker when the Laves phase was present only in a portion of the micropillar (Figure 5 d-f) and therefore the entire pillar was able to deform on a continuous basal slip plane in the Mg phase alone, even if this was further down the pillar. The small strengthening contribution was then presumably due to the slight radius increase along the pillar axis (Figure 5 f).

Similarly, the yield stress of micropillars oriented for extension twinning also increased due to the presence of the Laves phase ((Figure 5 g-i). Specifically, $\sigma_{0.2}$ increased from ~ 221 MPa for a pillar containing only α-Mg (Figure 5 h) to ~450 MPa for a pillar containing Laves phase and α-Mg phase (Figure 5 i).

Across all pillars tested, the effect of the geometrical alignment of the two phases therefore appears to be more important than the volume fraction of Laves phase in the micropillar with the complete obstruction of an easy deformation path inside the Mg matrix giving maximum strengthening and co-deformation with plasticity in the C36 phase or along an interface in the cases where the Mg phase cannot deform to give a through-thickness shape change of the pillar by itself.

It can be further seen that the Laves phase present in the micropillar is not homogenous and single phase, but that there are precipitates within the C36 Laves phase (Figure 1 (e)). These precipitates were also observed in undeformed micropillars indicating that they were not generated during deformation.

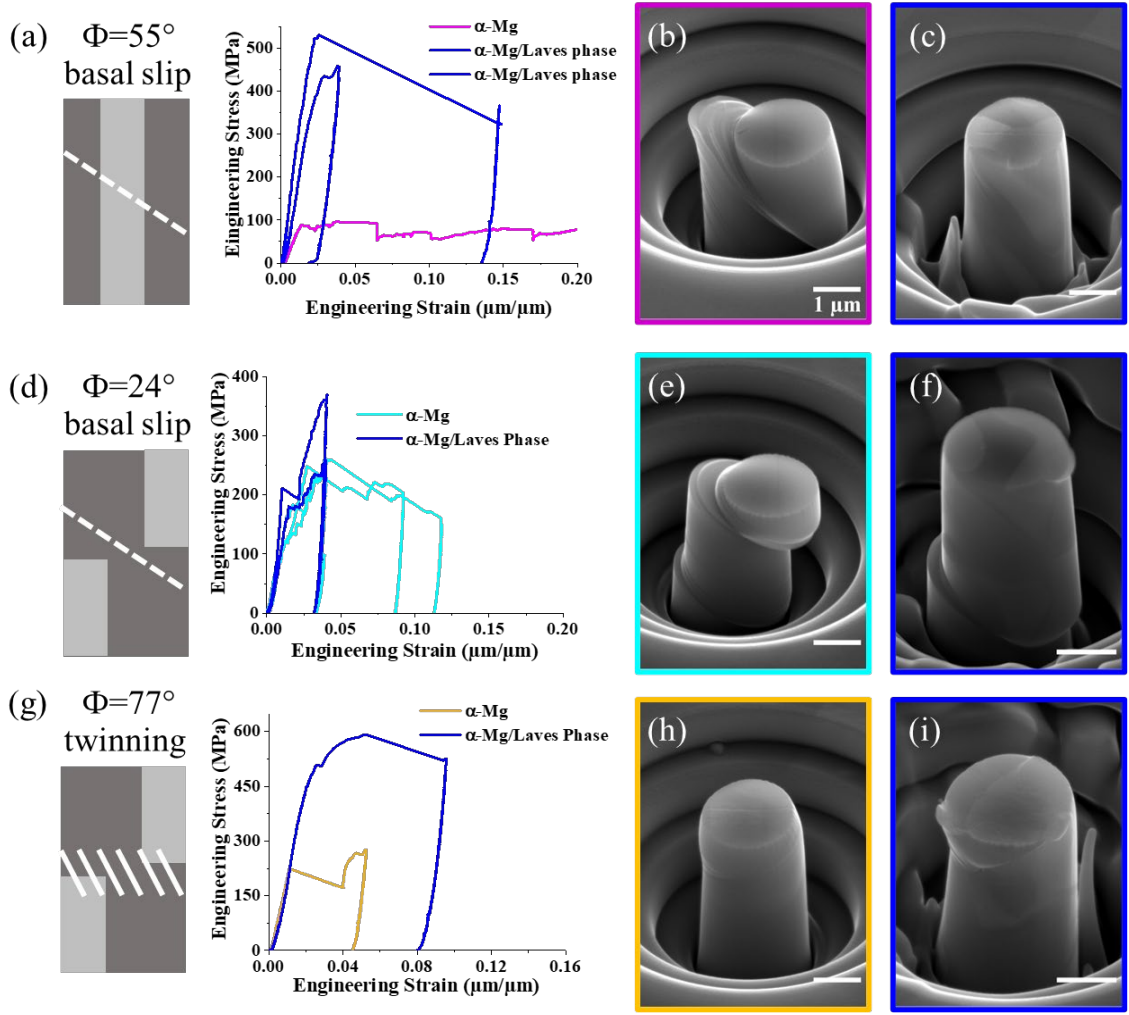

*Figure 5: Engineering stress-strain curves of Mg and Mg/C36 Laves phase micropillars with representative micrographs of corresponding micropillars. On the left, the three major cases depicted here are drawn schematically: (a-c) geometric alignment of Mg and C36 phase obstructing through-thickness deformation along the softer Mg phase in basal slip only; (d-f) deformation in case of continuous Mg deformation path by basal slip and (g-i) co-deformation in case of deformation dominated by extension twinning in the Mg matrix. The scale bar (length of white line) represents a length of 1 µm.*

### 3.4 Co-deformation of the α-Mg and cubic C15 CaAl₂ Laves phase

In contrast to the fine Mg/C36 composite in the as-cast material, the annealed state allowed the preparation of micropillars consisting entirely of either Mg or CaAl$_2$ Laves phase as well as a combination of the two. A direct comparison (Figure 6) between monolithic and two-phase pillars milled in adjacent grains of the two phases and at their interface, respectively, gave a much higher yield stress for the CaAl$_2$ phase (σ$_{0.2}$ ~2956 MPa) compared with the α-Mg matrix (σ$_{0.2}$ ~132 MPa) and an intermediate value for a pillar at the Mg/CaAl$_2$ interface.

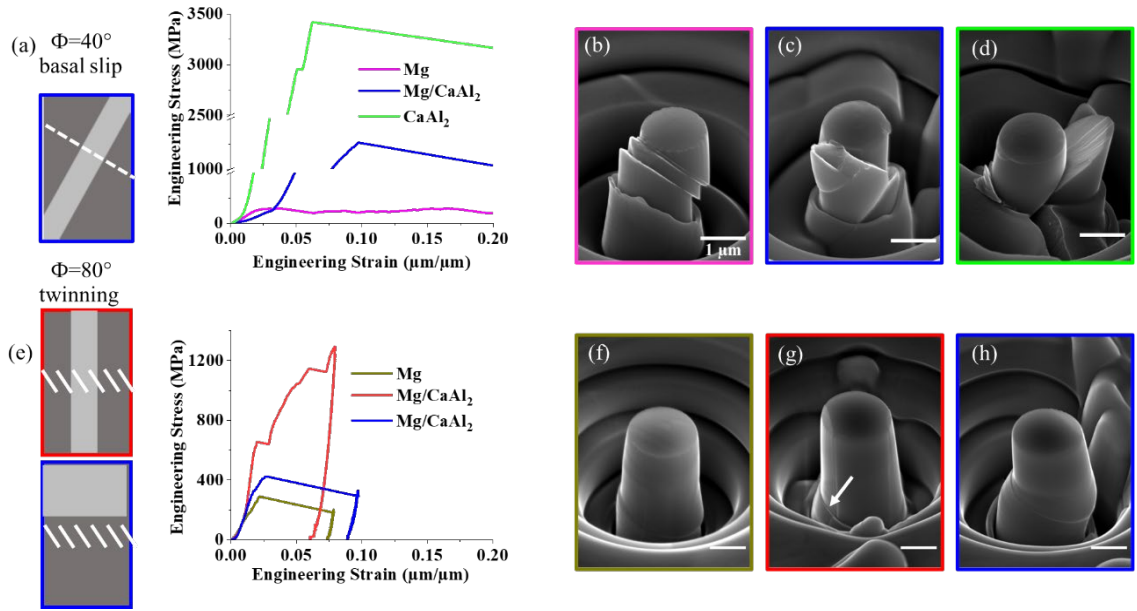

*Figure 6 (a): Engineering stress strain curves of micropillars oriented for basal slip containing α-Mg phase, α-Mg/CaAl$_2$ interface, and CaAl$_2$ phase. (b), (c), and (d) are the micrographs of micropillars containing α-Mg phase, α-Mg/CaAl$_2$, and CaAl$_2$ phase, respectively. Both (b) and (c) were milled into the same Mg grain which has high Schmid factor for basal slip. (e) Engineering stress strain curves of micropillars containing the α-Mg phase (ϕ = 80°) and a Mg/CaAl$_2$ interface (f), (g) and (h) are the SE images of the deformed micropillars. The frame colours of (f), (g) and (h) correspond to the respective stress-strain curves. The white arrow in (g) highlights the point of slip transfer from Mg to CaAl$_2$. Scale bar represent the length of 1μm.*

Only a single case of slip transfer visible at the micropillar surface was observed (Figure 6 (g)), although it could not be characterised in more detail due to its position at the very bottom of the micropillar. Again, the pillars showed very little strengthening where a continuous path through the Mg matrix existed (Figure 6 (h) and corresponding stress strain curve) and an intermediate value where the CaAl$_2$ phase could not be circumvented easily.

# 4 Discussion

## 4.1 Deformation of the α-Mg(Al, Ca) phase

We found that the Mg matrix phase contained about 1.2 % Al and a small amount of Ca below the limit of quantitative detection by EDS. This is consistent with expectations based on the solubility of the two elements [14]. In addition, we found precipitates of $CaAl_2$, presumably, parallel to the basal plane which have been described before [8, 11, 12]. In the light of these features, we set out to characterise the orientation dependence of flow in the α-Mg phase in terms of the dominant deformation mechanism (dislocation slip or twinning) and the associated critical resolved shear stresses.

We observed the expected anisotropy (see Figure 3 (a) and (b)) with the yield stress increasing as the inclination angle of the c-axis towards the compression direction tilts from 0° towards 90°. This is because the Schmid factor, $m_S$, for basal slip is reduced and higher stresses are required to reach the $\tau_{CRSS}$ for basal slip. Further, it can be seen in Figure 3 (a) that the pillar with an inclination angle of ϕ ~ 6 ° (black curve), exhibited no significant strain bursts in the stress-strain curve, but showed a rather continuous transition from the elastic to the plastic region and a higher strain hardening. The highest Schmid factor for basal slip in this pillar was ~0.09, but even then, some basal slip traces were observed on the surface of a deformed micro-pillar. The strong work hardening in such a pillar likely reflects simultaneous activation of non-basal (pyramidal) slip systems, which is usually associated with pillars milled in orientations with ϕ approaching 0° [43]. The $\tau_{CRSS}$ for basal slip was measured as 58 ± 19 MPa, in good agreement with the earlier reported values of ~ 51 MPa by Wang et al. [26] for micropillars of similar geometry and size in Mg5%Zn alloy. A more detailed comparison of the CRSS for basal slip with values reported in the literature is presented in Figure 7. These values are well above those observed in macroscopic tests [16, 17] due to the well-known size effect [40-42, 44-46]. The size effect occurs because of the small number of dislocations sources and the truncation of the moving dislocation arms associated with the small volume and/or with the scarcity of mobile dislocations [27, 40].

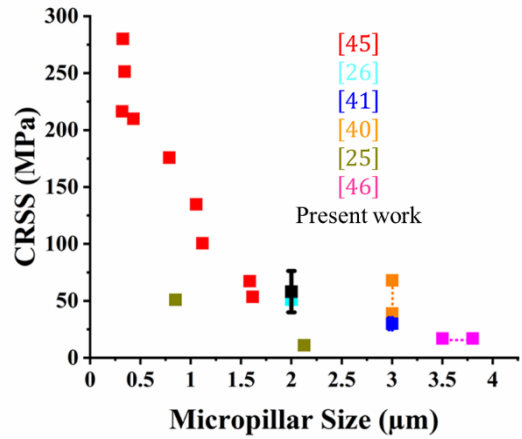

*Figure 7: Comparison of CRSS values for basal slip observed in this work and reported in existing literature.*

In the yield stress values of micropillars which were present in grains with ϕ ≥ 77 ° (Figure 3), i.e. away from those orientations where the measured yield stress can confidently be associated with basal slip, the experimental scatter increased significantly. The $\sigma_{0.2}$ for these orientations was 240 ± 90 MPa. A relatively large scatter for a similar orientation was also observed by Wang et al. [26] in Mg5%Zn and by Wang et al. [27] in a Mg-0.2Ca-1.8Zn alloy. They [27] related the observed scatter to the initial misalignment between the indenter and the micropillar surface as twin nucleation is very sensitive to localised stresses. We did not try to calculate the CRSS for twinning from the present data as the presence of a deformation twin can only be confirmed via lift out and additional characterisation by EBSD, TKD or TEM of each pillar. Nonetheless, we did confirm twinning by TKD of a cross-section of one micropillar and twin-like (lenticular) features were visible in many of the deformed micropillars oriented similarly for extension twinning. Extension twinning was also observed in similar orientations in micropillar compression of pure Mg and its alloys in several recent studies [27, 28, 43, 47-52]. In summary, the observations on the α-Mg micropillars compressed along different orientations conforms entirely with expectations for pillars of the given size and is consistent with reports on microcompression testing in other Mg alloys.

## 4.2 Deformation of the Laves phases

Laves phases (C15, C14 or C36) have a high strength compared to the α-Mg matrix. While to the best of our knowledge no data exists for the Ca(Mg,Al)$_2$ C36 phase alone, due to the difficulty of synthesising bulk specimens of sufficient size, we measured here a yield stress of nearly 3 GPa for the CaAl$_2$ phase. This value is within the same range as the values reported by Freund et al. [15] in their work on CaAl$_2$. Using micropillars of a similar diameter of ≈ 2 μm, the CRSS for basal slip in the CaMg$_2$ Laves phase has been measured as ≈ 520 MPa [4]. A consistent difference in the micro hardness of CaAl$_2$ and CaMg$_2$ was reported by Rokhlin et al [5]. The difference between the hardness of the CaMg$_2$ and CaAl$_2$ phases is evident in a comparison of work by Zehnder et al. [4] and Freund et al. [15]. Zehnder et al. [4] reported a hardness of ≈ 3.4 ± 0.2 GPa for CaMg$_2$, while Freund et al. [15] reported a hardness of 4.9 ± 0.7 GPa for the CaAl$_2$ phase, both at room temperature. We would expect the CRSS values for the C36 Laves phases to be within the range of values reported for the C14 and C15 Laves phases. In our case, we change the B element from pure Mg (C14) to a mixture of Mg and Al (C36) to pure Al (C15), however, in the binary Nb-Co system, all three Laves phases exist as NbCo$_2$ and no significant dependence on the crystal structure was found in indentation across a diffusion couple containing all three phases [53, 54].

In principle, we would therefore expect all three Laves phases to act as efficient strengthening phases and this is in fact what we observed in all micropillars containing both phases (α-Mg and Laves phase). Using the image of simple composite deformation, the micropillar strength is determined by the way the two phases are arranged geometrically, with an efficient strengthening only achieved where deformation along the 'weakest link' of a continuous soft phase with a suitable deformation path (basal slip in Mg) is prevented (Figure 5 - Figure 6).

## 4.3 (Co)-Deformation and reinforcement potential in bulk alloys

Following the characterisation of the Mg(Al, Ca) matrix phase's deformation and the co-deformation of pillars containing either C15 or C36 Laves phase, we focus here on the strengthening potential of these phases within a dual phase microstructure. In this context, the two most important aspects to consider are: (1) Can strengthening be achieved in such a way that the matrix is shielded from excessive stresses, but plastic co-deformation is enabled at stress concentrations to prevent early failure? (2) Can the desired phase be easily obtained in a morphology that allows effective stress transfer to the reinforcement phase?

To address the first question on whether plastic co-deformation is likely, we searched the surfaces of deformed micropillars from two-phase specimens for signs of co-deformation. In case of co-deformation of Mg and the C15 phase, no indications of plasticity traversing from the Mg into the Laves phase were found, with the one potential but unclear exception shown in Figure 6 (g). Several examples of two-phase pillars containing Mg and the C15 phase before and after deformation are shown in Figure 8. While significant plasticity was introduced overall in many cases, the strain appears to have been accommodated solely in the Mg phase and at the interfaces between Mg and the C15 Laves phase.

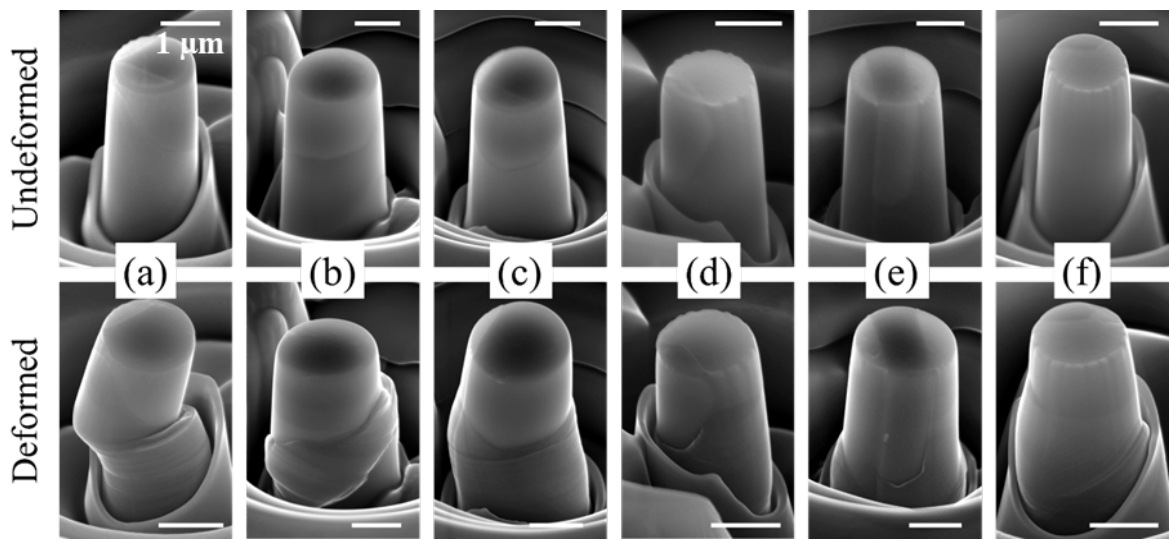

*Figure 8: Micropillars before and after deformation containing both the Mg and CaAl$_2$ C15 Laves phases. Deformation is constrained to the Mg phase and its interface with the Laves phase, but no indication of plasticity crossing the interface was found. The scale bars in the figure represent a length of 1μm.*

In a micropillars containing Mg and the hexagonal C36 phase, the post-mortem observations revealed slight differences (Figure 9). In this case, several pillars exhibited surfaces slip traces in both phases with again the majority of deformation achieved in the Mg phase, but clear indications of plasticity in the C36 phase as well. Four instances of plasticity in the C36 phase are shown at higher magnification in Figure 9 along with the micropillar viewed as a whole. In these cases, localised slip in the Mg phase appears to either have continued through smaller C36 phase particles (Figure 9 a and d) or the high stresses resulting from deformation of the surrounding volume led to the nucleation and/or motion of dislocation in the C36 phase directly, as may have been the case in the micropillars displayed in Figure 9 b and c.

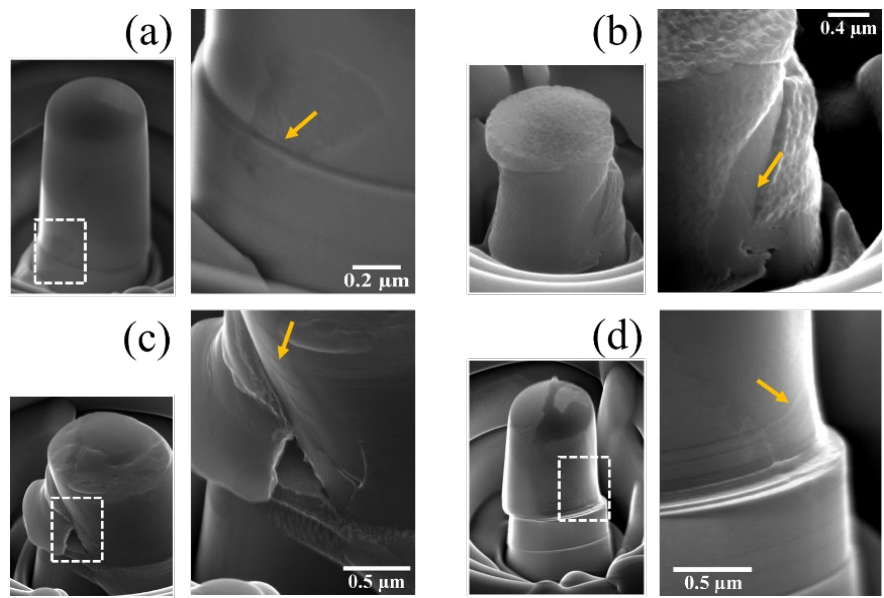

*Figure 9: Micropillars demonstrating co-deformation of α-Mg/C36 Laves phase micropillars. Orange arrows in (b), (c), (e) and (f) demonstrate the instances where slip transfer occurred into the Laves phase or/and where the Laves phase has undergone plastic deformation. (b) and (e) are magnified images of the areas bounded by blue rectangles in (a) and (d) respectively.*

As the phase distribution could not be exactly controlled, we cannot reach a definite conclusion based on this data, but it appears likely that co-deformation with the C15 phase is much more difficult, as otherwise similar morphologies (e.g. Figure 8 d-e and Figure 9 b-d) should also result in visible co-deformation.

While the C15 is harder, it also potentially offers a higher number of slip planes that may align with the dominant (basal) slip system in the Mg matrix. Indeed, an initially advantageous alignment of Mg(0002) // C15(111) has been identified for the initial stages of precipitation [14], but the interfaces from which we were able to cut two- and single phase micropillars (Figure 2) possess no common orientation relationship. For the C36 phase, we previously determined an orientation relationship ∠$[0001]_{Mg},[0001]_{C36}$ = 82 ° ± 10 ° and encountered several C36 particles with a c-plane parallel to Mg($1\bar{1}01$) in several TEM lamellae [6, 10, 14], while Luo et al. [34] found an orientation relationship with almost parallel basal planes between the two phases. The C36 Laves phase tends to deform via basal slip [6, 10] and is therefore able to accommodate plasticity on the Mg basal plane directly on its basal plane. Or following dislocation storage at the interface with the resulting non-basal dislocations in the Mg phase inducing basal slip in the Laves at a near 90° angle might occur, as proposed in [6] based on observations of plastic co-deformation in bulk composites with a C36 skeleton. The closely related C14 phase possesses a low CRSS for prismatic slip [4], but whether this is also an easy slip system in the C36 phase, with its changed stacking along the c-axis, is not known.

We therefore conclude that in spite of the potential of the $CaAl_2$ phase as a more isotropic cubic phase, its reinforcement potential for Mg-Al-Ca composite alloys is probably limited. In addition to the even more severely restricted co-deformation and therefore presumably limited bulk ductility, the $CaAl_2$ phase has not been observed in a useful morphology for load transfer. This is because the skeleton-like morphology is promoted with increasing Ca/Al ratio [8] and the $CaAl_2$ phase only appears at low Ca/Al ratios [30]. While the C36 and C14 phases form a skeleton in which strut thickness and connectivity can then be tuned by solidification rate and alloy content, respectively, the C15 phase precipitates nucleate as particles oriented along the basal plane, but these grow to globular particles. Neither characteristic, the initial alignment with the basal plane (rather than perpendicular to it to inhibit basal flow in the Mg matrix) nor the globular shape (allowing easy circumvention on parallel slip planes) provide a promising path to a microstructure in which the Mg matrix experiences shielding from an applied load. This is of particular importance under the creep conditions envisaged for these alloys, even if some (work) hardening has in fact been successfully achieved with these $CaAl_2$ microstructures [14, 55]. However, if a transformation path to C15 is found in which a skeleton reinforcement is achieved without or only slow further transformation to globular particles, then the phase would be a plausible candidate for reinforcement where phase stability to high temperatures well beyond 200 °C was of greater importance than the potential for limited co-deformation.

# 5 Conclusions

We investigated the deformation and co-deformation of a α-Mg(Al,Ca) solid solution matrix with small precipitates deformed along different crystal orientations and its co-deformation with the cubic $CaAl_2$ C15 and hexagonal $Ca(Mg,Al)_2$ C36 Laves phases in order to assess the relative potential of the two Laves phases for in-situ reinforcement of bulk alloys. Our findings and conclusions include:

1. The CRSS for basal slip in ≈ 2 µm α-Mg(Al,Ca) micropillars at 58 ± 19 MPa corresponds closely to that of related alloys and twinning is introduced where the compression direction approaches the orientation of the basal plane.
2. Both, the C36 and C15 Laves phases delay the onset of basal slip in the α-Mg phase and also provide significant strengthening where the Mg phase is oriented for extension twinning.
3. Although both Laves phases are clearly hard and brittle, the C36 phase showed several instances of clear co-deformation, while the C15 phase did not.
4. Overall, the C36 phase appears the more promising candidate for further alloy design in spite of its lower melting point, high temperature stability and hardness, as it offers superior co-deformation and a tuneable skeleton morphology to achieve load shielding under creep conditions for the Mg matrix.

# Acknowledgements

The authors gratefully acknowledge the financial support received from the Deutsche Forschungsgemeinschaft (DFG), under Collaborative Research Centre (CRC) 1394, project ID 409476157 within projects C01 and C02. We are also grateful to Prof. Hauke Springer and Dr. Leandro Tanure for providing the material (from project S in the same CRC) used in this study. We are also thankful to the Dr. Bengt Hallstedt, Dr. Siyuan Zhang, Dr. Wei Luo, Mr. Mattis Seehaus, Mr. Martin

Heller, and Mr. Lukas Berners for the fruitful discussions which turned out to be very helpful in this research.

# Data availability

Data will be made available on request.